\documentclass[amsmath,amssymb,twocolumn,prl]{revtex4}
\usepackage{wasysym}
\usepackage{graphicx}
\usepackage{float}

\def\al{\alpha}

\def\ga{\gamma}

\def\ze{\zeta}
\def\et{\eta}

\def\ka{\kappa}
\def\la{\lambda}

\def\ph{\phi}

\def\ch{\chi}
\def\ps{\psi}
\def\om{\omega}
\def\Ga{\Gamma}

\def\Om{\Omega}
\def\mn{{\mu\nu}}

\def\fr#1#2{{{#1} \over {#2}}}
\def\half{{\textstyle{1\over 2}}}

\def\frac#1#2{{\textstyle{{#1}\over {#2}}}}

\def\prt{\partial}

\def\etal{{\it et al.}}

\def\pt#1{\phantom{#1}}
\def\ol#1{\overline{#1}}

\def\ab{\overline{a}{}}

\def\cb{\overline{c}{}}

\def\eb{\overline{e}{}}

\def\sb{\overline{s}{}}

\def\afb{(\ab_{\rm{eff}})}

\def\afbx#1{(\ab^{#1}_{\rm{eff}})}

\def\afbw{\afbx{w}}

\def\axw{\afbw_{X}}
\def\ayw{\afbw_{Y}}
\def\azw{\afbw_{Z}}

\def\axw{\afbw_{X}}
\def\ayw{\afbw_{Y}}
\def\azw{\afbw_{Z}}

\def\cxxw{(\cb^w)_{XX}}
\def\cyyw{(\cb^w)_{YY}}

\def\cxyw{(\cb^w)_{(XY)}}
\def\cxzw{(\cb^w)_{(XZ)}}
\def\cyzw{(\cb^w)_{(YZ)}}
\def\ctxw{(\cb^w)_{(TX)}}
\def\ctyw{(\cb^w)_{(TY)}}
\def\ctzw{(\cb^w)_{(TZ)}}

\def\lrpartial{\raise 1pt\hbox{$\stackrel\leftrightarrow\partial$}}

\def\lrDmu{\stackrel{\leftrightarrow}{D_\mu}}

\def\vb#1#2{e_{#1}^{{\pt{#1}}#2}}
\def\ivb#1#2{e^{#1}_{{\pt{#1}}#2}}
\def\uvb#1#2{e^{#1#2}}

\def\mt{m^{\rm T}}
\def\ms{m^{\rm S}}

\newcommand{\beq}{\begin{equation}}
\newcommand{\eeq}{\end{equation}}
\newcommand{\bea}{\begin{eqnarray}}
\newcommand{\eea}{\end{eqnarray}}
\newcommand{\bit}{\begin{itemize}}
\newcommand{\eit}{\end{itemize}}
\newcommand{\rf}[1]{(\ref{#1})}

\begin{document}

\title{Superconducting-Gravimeter Tests of Local Lorentz Invariance}

\author{Natasha A.\ Flowers, Casey Goodge, Jay D.\ Tasson}

\affiliation{Physics and Astronomy Department, Carleton College, Northfield, MN 55057, USA}

\date{September 2017}

\begin{abstract}

Superconducting-gravimeter measurements
are used to test the local Lorentz invariance of the gravitational interaction
and of matter-gravity couplings.
The best laboratory sensitivities to date are achieved
via a maximum-reach analysis
for 13 Lorentz-violating operators,
with some improvements exceeding an order of magnitude.

\end{abstract}

\maketitle

Local Lorentz invariance
is among the foundational building blocks of General Relativity (GR).
Though GR provides an impressive description
of the wide variety of gravitational phenomena,
standard lore holds that 
GR may be the low-energy limit
of an underlying theory 
that merges gravitation and quantum physics,
such as string theory.
Local Lorentz violation may arise
in such an underlying framework \cite{ks}.
Hence tests of local Lorentz invariance
probe the core construction of GR
and may provide clues about the structure
of new physics at the quantum-gravity scale.
These ideas triggered the development of a
comprehensive effective field theory based framework \cite{SME,akgrav}
for testing Lorentz symmetry
used in many modern searches for violations \cite{data}.

Superconducting gravimeters \cite{sens}
have generated a vast amount of information about
the gravitational field of the Earth.
Devices functioning at over 2 dozen locations around the globe
generate data at minute intervals for the Global Geodynamics Project (GGP) \cite{ggp}.
In some cases
measurements span more than a decade,
and sensitivities to local variations in the gravitational field
approaching parts in $10^{12}$ can be extracted
for variations with periods on the order of a day.
Stability at the level of parts in $10^9$ per year \cite{sens}
has also been achieved.
Though the primary use of the data is in geophysical applications,
the nature of the data clearly also depends on the foundational theories of physics.
Hence these data sets provide
opportunities to test fundamental physics \cite{shiomi,gwav}.
The search for preferred frame effects in gravitational physics,
a particular Lorentz-symmetry violating scenario,
was perhaps the first application of superconducting gravimeters
to tests of foundational theory \cite{wg}.

In the 4 decades since those early tests,
interest in Lorentz violation has surged \cite{proc},
as have theoretical and experimental developments \cite{rev,revg,rev2}.
In addition to the search for preferred frame effects
as a signal of alternatives to GR \cite{will},
more general types of Lorentz violation are now actively sought
as a possible signal of new physics at the Planck scale \cite{data}.
Though performing Planck-scale experiments directly
will likely remain infeasible for the foreseeable future,
experimental information about the nature of the underlying theory
can be attained by searching for tiny Planck-suppressed effects
in experiments at presently accessible energies.
Lorentz violation provides a useful candidate Planck-suppressed effect \cite{ks},
and the gravitational Standard-Model Extension (SME) provides a field-theory based framework
for organizing a systematic search \cite{SME,akgrav,nonmin}.
While sensitivities to SME coefficients for Lorentz violation
have been achieved in a variety of gravitational systems 
\cite{hml,hmd,gexpt,llr,planetary,pulsar,akmmgw,cer},
including pioneering work with an atom-interferometer gravimeter \cite{hml,hmd},
this work
provides the first exploration of superconducting gravimeters
in the SME framework
and the first search for matter-sector Lorentz violation
using gravimeters of any kind.
Sensitivity improvements over prior gravimeter work \cite{hml,hmd} are achieved
for 7 coefficients for Lorentz violation,
and the best laboratory sensitivity to 6 coefficients 
not previously explored in gravimeter experiments
is achieved.
In some cases,
sensitivities are improved by more than a factor of 10.

The SME is constructed as an expansion about the actions of GR
and the Standard Model
in Lorentz-violating operators of increasing mass dimension.
In the present work we focus on the minimal gravitational SME,
in which attention is restricted to operators of mass dimension 3 and 4.
We consider both the pure-gravity sector \cite{lvpn} and the spin-independent
gravitationally-coupled fermion sector \cite{lvgap}
in the limit of linearized gravity.
Though work extending the framework 
to include higher dimension operators \cite{akmmgw,cer,ho}
and nonlinear gravity \cite{nlg}
is now well underway,
treatment of these operators lies beyond our present scope.
Here we summarize aspects of the SME framework relevant for this work.
For additional detail,
the reader is referred to Refs.\ \cite{akgrav,lvpn,lvgap}.

The SME action in this limit can be written
$ S = S_G + S_\ps + S^\prime$.
Here $S_G$ is the minimal pure-gravity sector,
\beq
S_G = \fr {1}{16 \pi G} \int d^4x e(R -u R 
+s^\mn R_\mn + t^{\ka\la\mu\nu} C_{\ka\la\mu\nu}),
\eeq
where 
$G$ is Newton's constant,
and $R$, $R_\mn$, and $C_{\ka\la\mu\nu}$ are the Ricci scalar, Ricci tensor, and Weyl tensor respectively.
The symbol
$e$ is the determinant of the vierbein $\vb \mu a$,
and
$u$, $s^\mn$, and $t^{\ka\la\mu\nu}$ are coefficient fields
having dynamics contained in $S^\prime$.
Lorentz violating signals in the post-newtonian analysis to follow
are associated with $s^\mn$,
without contribution from $t^{\ka\la\mu\nu}$ \cite{yb}.

Similarly spin-independent effects
in the minimal gravitationally-coupled fermion sector take the form
\beq
S_\ps= \int d^4 x \left[\half i e \ivb \mu a \ol \ps \Ga^a \lrDmu \ps 
- e \ol \ps M \ps \right].
\eeq
Here $\ps$ is the fermion field
and
$D_\mu$ is the covariant derivative,
which, along with the vierbein, provides the coupling to gravity,
$\Ga^a
\equiv 
\ga^a - c_{\mu\nu} \uvb \nu a \ivb \mu b \ga^b
- e_\mu \uvb \mu a$,
and 
$
M
\equiv 
m + a_\mu \ivb \mu a \ga^a$.
The matter-sector coefficient fields $a_\mu$, $e_\mu$, and $c_\mn$
also have dynamics contained within $S^\prime$.
The dynamics are assumed to trigger spontaneous Lorentz violation
in which the coefficient fields acquire vacuum expectation values,
a process for generating Lorentz violation in gravity
that is consistent with Riemann geometry \cite{akgrav,rbno}.
The issue of geometric consistency
has also led to the development of SME-based Finsler spacetimes
\cite{finsler}.

\begin{center}
\begin{tabular}{lc}
\multicolumn{2}{c} 
{Table I. Amplitudes for the force $F_{LV}$.} \\
\hline
\hline
Amplitude & Phase\\ 
\hline
$G^w_{\om} = 2 m^w \ze \cxzw$ & \\
$ \pt{G^w_{\om} =}
- \fr45 V_L \al \afbw_Y \sin \ch 
- 2 m^w V_L \ctyw \sin \ch$ &  
$\ph$ \\
$H^w_{\om} = 2 m^w \ze \cyzw $ & \\
$ \pt{H^w_{\om} =}
+ \fr{4}{5} V_L \al \afbw_X \sin \ch
+ 2 m^w V_L \ctxw \sin \ch$ & 
$\ph$ \\
$G^w_{2 \om} = m^w \ze (\cxxw - \cyyw)$ & 
$2 \ph $ \\
$H^w_{2 \om} =  2 m^w \ze \cxyw $ &
$2 \ph$ \\
$G^w_{\Om} = 2 V_\oplus \al (\ayw \cos \et + \azw \sin \et)$ & \\
$ \pt{G^w_{\Om} =}
+ 2 m^w V_\oplus \big[\ctyw \cos \et + 2 \ctzw \sin \et \big] $ &
$0$ \\
$H^w_{\Om} = - 2 V_\oplus \al \axw
- 2 m^w V_\oplus \ctxw $ &
$0$ \\
$E^{\prime w}_{\om} = - V_L \left( 2 \al \afbw_Y
+ \fr45 m^w \ctyw \right) \sin \ch$ &  
$\ph$ \\
$F^{\prime w}_{\om} = V_L \left( 2 \al \afbw_X
+ \fr45 m^w \ctxw \right) \sin \ch$ & 
$\ph$ \\
$E^{\prime w}_{\Om} = 2 V_\oplus \al (\ayw \cos \et + \azw \sin \et)$ & \\
$ \pt{E^w_{\Om} =}
+ 2 m^w V_\oplus (\ctyw \cos \et + \ctzw \sin \et) $ &
$0$ \\
$F^{\prime w}_{\Om} = - 2 V_\oplus \al \axw
- 2 m^w V_\oplus \ctxw $ &
$0$ \\
\hline
\hline
\end{tabular}
\end{center}

The vacuum expectation values,
or coefficients for Lorentz violation,
are denoted with an overline
and,
though other choices are possible \cite{cl},
are typically assumed constant in asymptotically Minkowski spacetimes.
For example,
the vacuum value associated with the coefficient field $c_\mn$
is $\cb_\mn$ satisfying $\prt_\al \cb_\mn = 0$.
The coefficients parameterize the amount of Lorentz violation in the theory
and are the objects sought by experiment.
Following generic treatment of spontaneous Lorentz violation
and the development of the post-newtonian limit
in the pure-gravity sector \cite{lvpn}
and the matter sector \cite{lvgap},
the signals for Lorentz violation in gravitational experiments can be found.
In the work to follow,
coefficients for Lorentz violation
$\ab_\mu$ and $\eb_\mu$ always appear in the combination $\afb = \ab_\mu - m \eb_\mu$.
Additionally,
$\afb_\mu$ appears with a constant $\al$ in gravitational studies,
which characterizes coupling constants in the underlying theory.
This combination is of special interest since it is typically unobservable
in flat spacetime \cite{akjt}.
Note also that the matter-sector coefficients
are in general particle-species dependent
and a superscript $w$ denotes the associated species.
In this work, 
the focus is on ordinary matter with $w$ referring to proton, neutron, or electron.

The system of interest here
can be referred to as a force-comparison gravimeter experiment \cite{lvgap}.
In this class of experiments,
the gravitational force on a laboratory body
is countered by an appropriate electromagnetic force,
and
the Lorentz-violating signal
can be written
\beq
F_{LV} =
- \mt g \sum_n \left(
A_n \cos (\om_n T + \ph_n) +
B_n \sin (\om_n T + \ph_n)  
\right),
\label{Fz}
\eeq
as developed in Sec.\ VIIC of Ref.\ \cite{lvgap}, where 
\bea
A_n &=& \sum_w \left (\fr{N^w}{\mt} G_n^w 
+ \fr{N^w_\oplus}{\ms} E_n^{\prime w} 
+ \frac{1}{3} G_n \right),
\nonumber \\
B_n &=& \sum_w \left(\fr{N^w}{\mt} H_n^w +
 \fr{N^w_\oplus}{\ms} F_n^{\prime w} 
+ \frac{1}{3} H_n \right).
\label{smeamp}
\eea
Here $g$ is the newtonian gravitational field,
$\mt$ and $\ms$ are the conventional Lorentz-invariant mass of the test body
and source body respectively, and
$N^w$ and $N^w_\oplus$ are the number of particles of type $w$
in the test body and the Earth respectively.
Here the test body is a niobium sphere with a mass of a few grams,
and the source body is the Earth.
The sum over $w$ runs over proton, neutron, and electron.
The frequencies $\om_n$ are drawn from the set
\beq
 \omega_n \in \{2 \om, \om, 2 \om + \Omega, 2 \om - \Omega, \om + \Omega, \om-\Omega,\Omega\},
\eeq
where $\om$ is the sidereal angular frequency
and $\Om$ is the annual angular frequency.
Note that $2\om$ arises due to the rotation
of 2-index coefficients.
The corresponding phase $\ph_n$ can be obtained from the frequency via the replacement
$\om \rightarrow \ph$, $\Om\rightarrow 0$,
where $\ph$ is a phase that specifies the orientation of the Lab at time $T=0$.
The time $T$,
along with the spacial coordinates $X,Y,Z$ are the coordinates of the 
Sun-centered celestial equatorial frame in standard use for SME studies \cite{data}.
The contributions to the Lorentz-violating amplitude
$G_n^w$, $H_n^w$, $E_n^{\prime w}$, and $F_n^{\prime w}$ can be found in Table I,
while the contributions $G_n$ and $H_n$ are constructed via Eq.\ (142) of Ref.\ \cite{lvgap}.
These are the results developed in \cite{lvgap}
presented here with a few corrections.
Here $V_L=\om R$, where $R$ is Earth's radius and
$V_\oplus$ is the speed of the Earth on its path around the Sun.
The angle $\ze$ is between the local Lorentz-invariant free-fall direction
and the direction of Earth's center,
$\ch$ is the colatitude of the experiment,
$\et$ is the inclination of Earth's orbit,
and $m^w$ is the mass of species $w$.

Our method for extracting measurements of the coefficients for Lorentz violation
from the GGP data
proceeds as follows.
We use corrected minute data,
which provides a measurement of the gravitational force
each minute obtained from the raw data
via some repairs performed by the station manager
including the removal of some transients such as major earthquakes.
Where possible, 
we follow the methods developed for the atom-interferometer gravimeter analysis \cite{hml}.
The basic idea is to perform a discrete Fourier transform on relevant sets of gravitational force
vs.\ time data
to extract the amplitudes $A_n,B_n$.
Equation \rf{smeamp} is then used to interpret the
amplitudes as measurements of the SME coefficients.

As is typical of SME searches,
the amplitudes $A_n,B_n$ extracted from data collected by a particular device at a given site
provide a measurement of a linear combination of SME coefficients
rather than a measurement of a single term in the underlying theory.
The numbers multiplying the coefficients for Lorentz violation
in these linear combinations can contain the colatitude of the experiment $\ch$
and dependence on the particle species content of the bodies involved.
Hence different sets of data from different locations and/or different devices
measure different linear combinations of SME coefficients.
Two procedures are common in the literature for extracting sensitivities
to individual SME coefficients from such linear combinations.
One approach effectively considers a series of special models,
each involving one and only one nonzero coefficient for Lorentz violation,
hence attributing the measurement of an amplitude $A_n,B_n$
to each of the coefficients it contains individually.
This approach is motivated by the thinking that exact cancellation
between multiple coefficients in a given measurement is unlikely.
We call sensitivities to coefficients for Lorentz violation
obtained in this way the ``maximum reach'' achieved for the given coefficient.
The other approach is to treat all coefficients as nonzero simultaneously
and use multiple sets of data to separate the linear combinations.
We say sensitivities attained in this way are found by ``coefficient separation''.
In what follows we apply both methods
using several sets of gravimeter data.

\begin{figure}[H]
\centering
\includegraphics[trim={0 1cm 0 2cm},clip,width=\columnwidth]{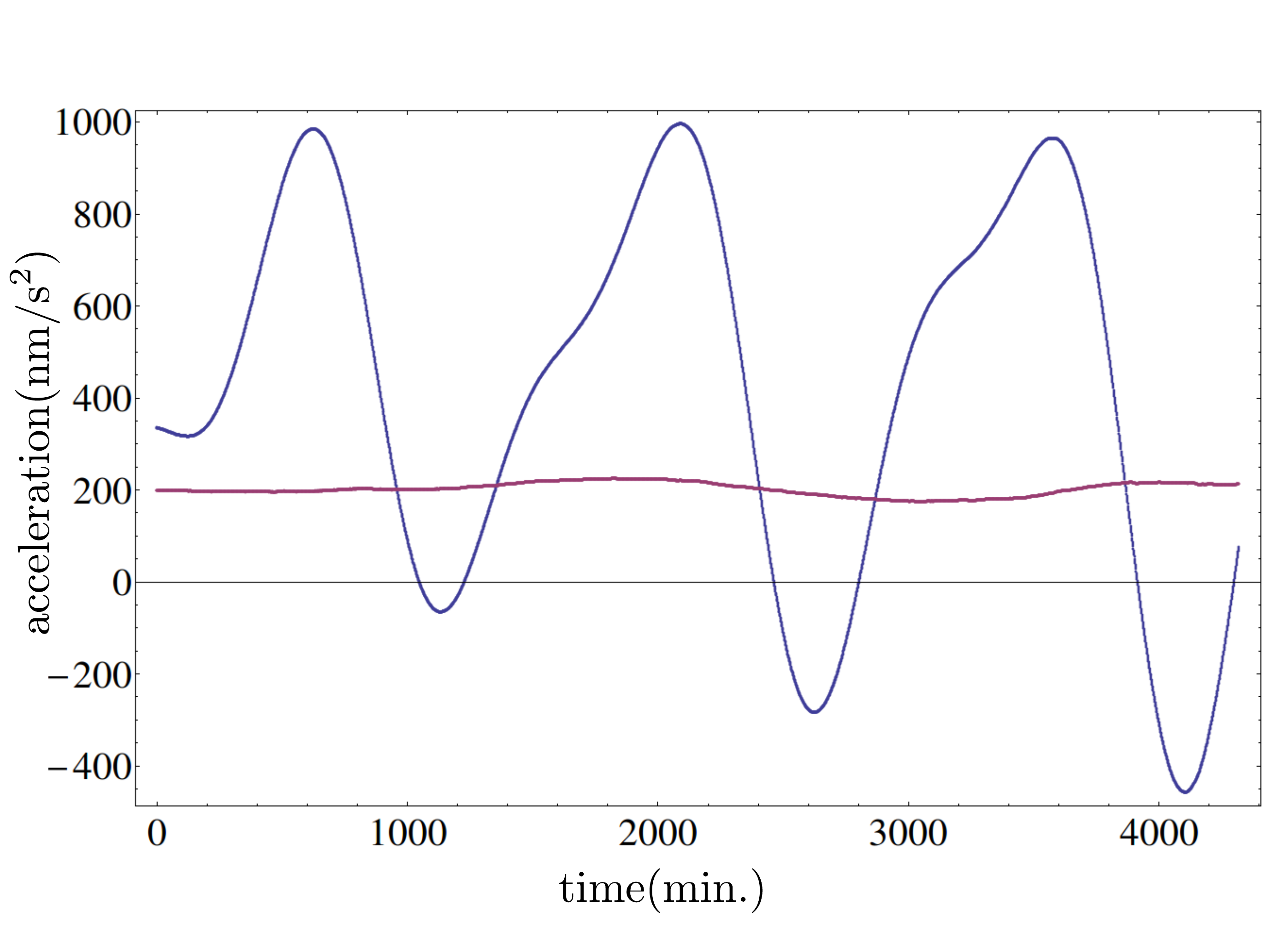}
\caption{\label{fig:data} Bad Homburg data taken
Jan.\ 1-3, 2012, before and after tidal model subtraction.
Discrete points are plotted
that appear as a continuous curve at this scale.}
\end{figure}

For our maximum-reach analysis 
we used data from Bad Homburg, Germany, from 2007-2013
(with several gaps of less than 1 week),
a site providing some of the cleanest data.
Three days of original gravimeter data from Bad Homburg is shown in Fig.\ \ref{fig:data},
appearing as the large-amplitude signal.
A daily variation associated with tidal effects
is clearly visible in the three peaks. 
Following Ref.\ \cite{hml}
we remove the dominate tidal contributions from the signal
using a model of solid Earth tides \cite{tide}.
Figure \ref{fig:data} also shows the data
after the subtraction of the model.
Application of discrete Fourier transform
to the residuals,
\beq
A_n,B_n = \frac{2}{K} \sum\limits_k d(t_k) \cos,\sin(\omega_n t_k + \phi), 
\eeq
yields the amplitudes shown in Table II.
Here, $K$ is the total number of measurements, 
$d(t_k)$ are the residual gravity measurements at times $t_k$.
Estimated uncertainties are obtained following Refs.\ \cite{hml,hmd}
by performing the analysis at several frequencies 
near the characteristic frequencies and computing 
the root mean square. 

\begin{center}
Table II. Bad Homburg amplitudes.\\
\begin{tabular}{c|c||c|c}
\hline
\hline
   Amp.                        & Meas. $(10^{-9} g)$  & Amp.                        & Meas. $(10^{-9} g)$\\
\hline
$ A_{2 \om} $                     & $-0.02 \pm 0.01 $ & $ B_{2 \om} $    & $\pt{-}0.04 \pm 0.01 $ \\   
$ A_{ \om} $                       & $-0.01 \pm 0.06 $ &  $ B_{ \om} $    & $-0.1 \pm 0.1 $\\          
$ A_{2 \om + \Omega} $              & $-0.003 \pm 0.004 $ &     $ B_{2 \om + \Omega} $   & $\pt{-}0.003 \pm 0.004 $ \\    
$ A_{2 \om - \Omega} $             & $-0.01 \pm 0.01 $ &    $ B_{2 \om - \Omega} $   & $\pt{-}0.006 \pm 0.005 $ \\ 
$ A_{ \om + \Omega} $              & $-0.00 \pm 0.02 $ &    $ B_{ \om + \Omega} $  & $-0.01 \pm 0.02 $  \\ 
$ A_{ \om - \Omega} $              & $\pt{-}0.01 \pm 0.03 $ &  $ B_{ \om - \Omega} $  & $\pt{-}0.06 \pm 0.03 $ \\    
$ A_{ \Omega} $                         & $-1 \pm 1 $ & $ B_{ \Omega} $           & $\pt{-}1 \pm 1 $ \\   
\end{tabular}
\end{center}

The amplitudes in Table II together with 
the maximum-reach procedure yield the sensitivities
to the coefficients for Lorentz violation shown in 
the second column of Table III.
A dagger $\dagger$ indicates a sensitivity
that exceeds previous laboratory tests,
though
better constraints exist from Solar System
or astrophysical observations
\cite{llr,planetary,pulsar,cer,bcer}.
The maximum reach listed here for the $\sb^\mn$ coefficients,
which have previously been explored via gravimeter analysis \cite{hml,hmd},
is an improvement upon that work for all 7 coefficients listed.

We perform the same analysis on data collected
from the device in Metsahovi, Finland, from 2007-2012,
and on a year's worth of data from Strasbourg, France
and from Apache Point, USA
taken in 2012.
While the maximum reach available from these sites
is typically less than that obtained from Bad Homburg,
their locations at different colatitudes
permit some coefficient separation.
We do this following the procedure outlined in Ref.\ \cite{hmd}
in which each measurement of $A_n,B_n$
provides a probability distribution that we assume to be Gaussian
with the measurement and uncertainty providing
the center and standard deviation.
The probability distribution can then be understood
as a function of the coefficients for Lorentz violation
through Eq.\ \rf{smeamp}.
These probability distributions can then be multiplied
together for each of the relevant measurements
from each of the 4 sites to obtain an overall probability distribution.
Integrating the distribution
over all of the coefficients except the one of interest
then yields an estimate and uncertainty for that coefficient.
The result of this process provides
our estimates for the coefficients achieved by coefficient separation
shown in the right column of Table III.
This procedure for achieving initial coefficient separation estimates
assumes the error sources in the 4 experiments
are completely independent,
while some geophysical noise sources
may be somewhat coherent.
Though beyond our current scope,
it may be possible to address this potential issue
through coherent combination of the original data.
Relative to Ref.\ \cite{hmd},
correlations between amplitudes due to finite data
are small and are neglected here.

\begin{center}
Table III.  Lorentz violation measurements. 
\begin{tabular}{c|rl|rl}
\hline
\hline
 Coeff.                     & & Max reach via & & 4-site Coeff. \\
                     & & Bad Homburg   & & separation \\
\hline
$ \bar s^{XX-YY} $      &  $2 $ & \hspace{-.05 in}$ \pm 1 \times 10^{-10\dagger}$ & & \hspace{.2 in} - \\ 
$ \bar s^{XY} $        &  $-4 $ & \hspace{-.05 in}$ \pm 1 \times 10^{-10\dagger}$ & & \hspace{.2 in} - \\   
$ \bar s^{XZ} $        &  $0 $ & \hspace{-.05 in}$ \pm 1 \times 10^{-10\dagger}$ &  $-2 $ & \hspace{-.05 in}$ \pm 2 \times 10^{-9\dagger}$\\   
$ \bar s^{YZ} $        &  $3 $ & \hspace{-.05 in}$ \pm 1 \times 10^{-10\dagger}$ &  $4 $ & \hspace{-.05 in}$ \pm 3 \times 10^{-9\dagger}$\\
$ \bar s^{TX} $        &  $-3 $ & \hspace{-.05 in}$ \pm 3 \times 10^{-7\dagger}$  &  $-3 $ & \hspace{-.05 in}$ \pm 3 \times 10^{-7\dagger}$ \\
$ \bar s^{TY} $        &  $-6 $ & \hspace{-.05 in}$ \pm 3 \times 10^{-7\dagger}$  &  $-5 $ & \hspace{-.05 in}$ \pm 2 \times 10^{-7\dagger}$ \\   
$ \bar s^{TZ} $        &  $-1 $ & \hspace{-.05 in}$ \pm 1 \times 10^{-6\dagger}$  &  $-1 $ & \hspace{-.05 in}$ \pm 1 \times 10^{-6\dagger}$ \\
$(\bar c^n) _{(TX)} $    &  $-4 $ & \hspace{-.05 in}$ \pm 6 \times 10^{-6}$ &  $-3 $ & \hspace{-.05 in}$ \pm 2 \times 10^{-3}$  \\
$(\bar c^n) _{(TY)} $    &  $-1 $ & \hspace{-.05 in}$ \pm 1 \times 10^{-5}$ &  $2 $ & \hspace{-.05 in}$ \pm 4 \times 10^{-3}$  \\
$(\bar c^n) _{(TZ)} $    &  $-1 $ & \hspace{-.05 in}$ \pm 1 \times 10^{-5}$ & & \hspace{.2 in} - \\
$ \alpha (\bar a^{e+p} _{\rm eff})_X $
         &   $-4 $ & \hspace{-.05 in}$ \pm 6 \times 10^{-6}$ \; \rm  GeV$^\dagger$
         &   $3 $ & \hspace{-.05 in}$ \pm 2 \times 10^{-2}$ \; \rm  GeV\\
$ \alpha (\bar a^n _{\rm eff})_X $  
         &  $-4 $ & \hspace{-.05 in}$ \pm 5 \times 10^{-6}$ \; \rm  GeV$^\dagger$
         &  $-3 $ & \hspace{-.05 in}$ \pm 2 \times 10^{-2}$ \; \rm  GeV\\
$ \alpha (\bar a^{e+p} _{\rm eff})_Y $  
         &  $-5 $ & \hspace{-.05 in}$ \pm 7 \times 10^{-6}$ \; \rm  GeV$^\dagger$
         &  $0 $ & \hspace{-.05 in}$ \pm 4 \times 10^{-2}$ \; \rm  GeV\\
$ \alpha (\bar a^n _{\rm eff})_Y $  
         &  $-4 $ & \hspace{-.05 in}$ \pm 6 \times 10^{-6}$ \; \rm  GeV$^\dagger$
         &  $0 $ & \hspace{-.05 in}$ \pm 3 \times 10^{-2}$ \; \rm  GeV\\
$ \alpha (\bar a^{e+p} _{\rm eff})_Z $  
          &  $-1 $ & \hspace{-.05 in}$ \pm 2 \times 10^{-5}$ \; \rm  GeV$^\dagger$  & & \hspace{.2 in} -\\
$ \alpha (\bar a^n _{\rm eff})_Z $  
          &  $-1 $ & \hspace{-.05 in}$ \pm 1 \times 10^{-5}$ \; \rm  GeV$^\dagger$ & & \hspace{.2 in} -\\
\end{tabular}
\end{center}

Coefficients $\bar s^{XX-YY}$ and $\bar s^{XY}$ are obtained
from amplitudes in which they are the only coefficient for Lorentz violation
involved.  
Hence these entries in the maximum reach column of Table III could equally be regarded
as the results of coefficient separation,
and they are omitted from the 4-site analysis.
Sufficient information is not available
to separate the $\alpha (\bar a^{e+p} _{\rm eff})_Z$,
$\alpha (\bar a^{n} _{\rm eff})_Z$, and $(\bar c^n)_{TZ}$ coefficients from each other.
Hence individual constraints are not available for column 3 of Table III,
and the combination is treated as a single coefficient
in the separation analysis resulting in
$\alpha (\bar a^{e+p} _{\rm eff})_Z 
+ 1.1 \alpha (\bar a^{n} _{\rm eff})_Z 
+1.1 {\rm GeV} \,\, (\bar c^n)_{TZ}
= 0 \pm 6 \times 10^{-4}$ GeV.
We do not include data from other experiments
beyond the 4 gravimeter sites
except in excluding from consideration other coefficients
that have been constrained much more tightly by nongravitational tests.
As this work was completed,
the $(\bar c^n)_{TJ}$ were also constrained by nongravitational tests \cite{cng}.
Note that the results of coefficient separation
generate improvements over prior lab work for $\sb^\mn$ coefficients
while constraints for matter-sector coefficients
are weak.
This feature can be traced to the fact that all 4 sites involve
niobium test masses and the Earth
and hence the same proton/neutron ratios.
Note also that proton and electron coefficients are listed together
as separating them would require charged matter.

The sensitivities to coefficients in Table III,
found via the standard approach to gravimeter analysis in the SME  \cite{hml,hmd},
provide a basic sense of upper bounds on coefficients.
However,
some care should be used in interpreting the results.
Though we find no compelling evidence of Lorentz violation,
some notable deviations from zero are seen
in a few cases.
In addition to the statistical expectation of a few weak signals
when seeking this number of effects,
these likely reflect some challenges inherent to the search
that we outline here.

The search involves subtracting dominant tidal effects
from the gravimeter signal
and attributing any remaining periodicity 
at the characteristic frequencies to Lorentz violation,
with uncertainty estimated by the average level of the local Fourier spectrum
near the characteristic frequency.
The method relies on the assumption that any potential Lorentz-violating signal
is not also contained in the tidal model.
Modeling of additional local effects
is avoided to minimize this concern.
We also note that a Fourier transform of the raw data
with no tidal modeling
yields the same level of reach for annual variations,
which is the aspect of the measurement associated with
many of the most significant sensitivity improvements.
The method also assumes that residual environmental effects
at the characteristic frequencies have a size similar
to neighboring Fourier amplitudes.
This assumption is most challenged by
Lorentz-violating frequencies that coincide
with dominant tidal components.
Here the relatively small residual signal is the result of subtracting
a comparatively large modeled tide from a similarly sized signal.
One could also imagine the Lorentz-violation signal
of a special linear combination of coefficients
that matches the tidal phase
being masked by a tidal effect.

A variety of opportunities for further improvements
with related experiments exist.
One key challenge in gravimeter tests
is managing periodic environmental effects
without using models constructed by fitting to gravimeter data.
A way to side-step this issue 
for matter-sector coefficients is to consider analogous Weak-Equivalence Principle tests
that search for variation in the relative gravitational force or acceleration of 2 or more bodies.
It may also be possible to use the phase information
associated with environmental systematics
to separate them from the effects of certain combinations
of coefficients for Lorentz violation.
Correlations between signals at multiple sites may also be useful.
Gravimeter data involving bodies of other compositions would aid 
in performing coefficient separation for the matter sector.
Free-fall gravimeter tests such as atom interferometers are also of interest,
particularly for the matter sector,
as they involve different dependence on the matter-sector coefficients.
An increase in the long-term stability of gravimeters would further improve
sensitivities at the annual frequency.
Beyond gravimeters,
searches for Lorentz violation with satellite geodesy data may be of interest.
In all,
exciting prospects remain for further searches for Lorentz violation
with gravimeters and related systems.

This work was supported in part 
by the Carleton College Clinton Ford Physics
Research Fund.

\end{document}